\documentclass[12pt]{article}
\usepackage{amsmath}
\usepackage{amssymb}

\input{tcilatex}

\begin{document}

\begin{center}
\smallskip \ 

\textbf{SL(2,R)-GEOMETRIC PHASE SPACE}

\smallskip \ 

\textbf{AND (2+2)-DIMENSIONS}

\smallskip \ 

R. Flores$^{\ast }$ \footnote{%
rflorese@gauss.mat.uson.mx}, J. A. Nieto$^{\star \ast }$ \footnote{%
niet@uas.edu.mx, janieto1@asu.edu}, J. Tellez$^{\ast }$ \footnote{%
jtellez@cajeme.cifus.uson.mx},

\smallskip \ 

E. A. Leon$^{\star }$ \footnote{%
ealeon@uas.edu.mx}, E. R. Estrada$^{\dagger }$ \footnote{%
profe.emmanuel@gmail.com}

\smallskip \ 

$^{\ast }$\textit{Departamento de Investigaci\'{o}n en F\'{\i}sica de la
Universidad de Sonora, 83000, Hermosillo Sonora, M\'{e}xico}

\smallskip

$^{\star }$\textit{Facultad de Ciencias F\'{\i}sico-Matem\'{a}ticas de la
Universidad Aut\'{o}noma} \textit{de Sinaloa, 80010, Culiac\'{a}n Sinaloa, M%
\'{e}xico.}

\smallskip

$^{\dagger }$\textit{Instituto Tecnol\'{o}gico Superior de Eldorado,
Eldorado, Sinaloa, Mexico}

\smallskip

\textbf{Abstract}
\end{center}

We propose an alternative geometric mathematical structure for arbitrary
phase space. The main guide in our approach is the hidden SL(2,R)-symmetry
which acts on the phase space changing coordinates by momenta and \textit{%
vice versa}. We show that the SL(2,R)-symmetry is implicit in any symplectic
structure. We also prove that in any sensible physical theory based on the
SL(2,R)-symmetry the signature of the flat target \textquotedblleft
spacetime\textquotedblright \ must be associated with either one-time and
one-space or at least two-time and two-space coordinates. We discuss the
consequences as well as possible applications of our approach on different
physical scenarios.

\smallskip \ 


Pacs numbers: 04.20.Gz, 04.60.-Ds, 11.30.Ly

April, 2013

\newpage

\noindent \textbf{1. Introduction}

\smallskip \ 

\noindent The importance of the $SL(2,R)$-group in physics and mathematics,
specially in string theory [1], two dimensional black holes [2] and
conformal field theory [3]-[4], has been recognized for long time. Recently
such a group structure has been considered as the key structure in the
development of two-time physics (2t-physics) (see [5]-[7] and references
therein). An interesting aspect is the relevance of the $SL(2,R)$-group in
2t-physics emerging from the Hamiltonian formalism of ordinary classical
mechanics. In fact, the $SL(2,R)$-group acts on a phase space, rotating
coordinates by momenta and \textit{vice versa}. Requiring this symmetry for
the constraint Hamiltonian system leads us to the conclusion that the flat
target \textquotedblleft spacetime\textquotedblright \ must have either a
(1+1)-signature or at least a (2+2)-signature [8]. However, this result
still requires a refined mathematical proof.

Specifically, we prove, in two alternative ways, that in a constraint
Hamiltonian formalism, in which the groups $SL(2,R)$ and $SO(t,s)$ are
symmetries of a classical system, the possible values for $t$-time and $s$%
-space are $t=1$ and $s=1$ or $t\geqq 2$ and $s\geqq 2$. In the process, we
formalize an alternative geometric structure for the phase space based on
the $SL(2,R)$-group.

As an application of our formalism, we develop the Dirac type equation in $%
(2+2)$-dimensions. We show that the $SL(2,R)$-group is relevant to
understand such equation.

The structure of this paper is as follows. In sections 2 and 3, we develop
the necessary steps to highlight the importance of the $SL(2,R)$-group in
classical constraint Hamiltonian systems. In section 4, we prove the main
proposition mentioned above. In section 5, we construct the Dirac type
equation in $(2+2)$-dimensions. Finally, in section 6 we make some
additional comments.

\bigskip \ 

\noindent \textbf{2. Lagrange-Hamiltonian system}

\smallskip \ 

\noindent Let us consider the action

\begin{equation}
S[q]=\int dtL(q,\dot{q}),  \tag{1}
\end{equation}%
where the Lagrangian $L=L(q,\dot{q})$ is a function of the $q^{i}$%
-coordinates and the corresponding velocities $\dot{q}^{i}\equiv dq^{i}/dt,$
with $i,j=1,\dots ,n$.

The canonical momentum $p_{i}$ conjugate to $q^{i}$ is defined to be

\begin{equation}
p_{i}\equiv \frac{\partial L}{\partial \dot{q}^{i}},  \tag{2}
\end{equation}%
Thus the action (1) can be rewritten in the form

\begin{equation}
S[q,p]=\int dt(\dot{q}^{i}p_{i}-H_{c}),  \tag{3}
\end{equation}%
where $H_{c}=H_{c}(q,p)$ is the canonical Hamiltonian,

\begin{equation}
H_{c}(q,p)\equiv \dot{q}^{i}p_{i}-L.  \tag{4}
\end{equation}

If one considers $m$ first class Hamiltonian constraints $H_{A}(q,p)\approx
0 $ (here the symbol "$\approx $" means weakly equal to zero [9]-[11]), with 
$A=1,2...,m$, then the action (3) can be generalized as follows:

\begin{equation}
S[q,p]=\int dt(\dot{q}^{i}p_{i}-H_{c}-\lambda ^{A}H_{A}).  \tag{5}
\end{equation}%
Here, $\lambda ^{A}$ are arbitrary Lagrange multipliers.

The Poisson bracket for arbitrary functions $f(q,p)$ and $g(q,p)$ of the
canonical variables $q$ and $p$ is defined as usual

\begin{equation}
\{f,g\}=\frac{\partial f}{\partial q^{i}}\frac{\partial g}{\partial p_{i}}-%
\frac{\partial f}{\partial p_{i}}\frac{\partial g}{\partial q^{i}}.  \tag{6}
\end{equation}%
Using (6) we find that

\begin{equation}
\begin{array}{c}
\{q^{i},q^{j}\}=0, \\ 
\\ 
\{q^{i},p_{j}\}=\delta _{j}^{i}, \\ 
\\ 
\{p_{i},p_{j}\}=0,%
\end{array}
\tag{7}
\end{equation}%
where the symbol $\delta _{j}^{i}$ denotes a Kronecker delta.

\bigskip \ 

\noindent \textbf{3. }$SL(2,R)$\textbf{-Hamiltonian system}

\smallskip \ 

\noindent It turns out that an alternative possibility to analyze the
previous program has emerged in the context of 2t-physics [5]-[7] (see also
Ref. [12] and references therein). The key point in this new approach is the
realization that, since in the action (3) there is a hidden invariance $%
SL(2,R)\sim Sp(2,R)\sim SU(1,1)$, one may work in a unified canonical phase
space of coordinates and momenta. Let us recall how such a hidden invariance
emerges. Consider first the change of notation%
\begin{equation}
q_{1}^{i}\equiv q^{i},  \tag{8}
\end{equation}%
and

\begin{equation}
q_{2}^{i}\equiv p^{i}.  \tag{9}
\end{equation}%
These two expressions can be unified by introducing the object $q_{a}^{i}$,
with $a=1,2$. The next step is to rewrite (5) in terms of $q_{a}^{i}$ rather
than in terms of $q^{i}$ and $p^{i}$. One finds that, up to a total
derivative, the action (5) becomes [4] (see also Refs. [12] and [13])

\begin{equation}
S=\int_{t_{i}}^{t_{f}}dt\left( \frac{1}{2}J^{ab}\dot{q}%
_{a}^{i}q_{bi}-H(q_{a}^{i})\right) .  \tag{10}
\end{equation}%
Here, $J^{ab}=-J^{ba}$, where $J^{12}=1$ is the antisymmetric $SL(2,R)$%
-invariant density (some times denoted with the symbol $\varepsilon ^{ab}$)
and 
\begin{equation}
H(q_{a}^{i})=H_{c}+\lambda ^{A}H_{A}.  \tag{11}
\end{equation}%
According to Dirac's terminology in the constrained Hamiltonian systems
formalism [14] (see also Refs. [9]-[11]), (11) corresponds to a total
Hamiltonian. From the action (10) one observes that, while the $SL(2,R)$%
-symmetry is hidden in (5), now in the first term of (10) it is manifest.
Thus, it is natural to require the same $SL(2,R)$-symmetry for the total
Hamiltonian $H(q_{a}^{i})$.

Consider the usual Hamiltonian for a free non-relativistic point particle

\begin{equation}
H=\frac{p^{i}p^{j}\delta _{ij}}{2m}+V(q),  \tag{12}
\end{equation}%
with $i=\{1,2,3\}$. According to the notation (8)-(9) we have

\begin{equation}
H=\frac{q_{2}^{i}q_{2}^{j}\delta _{ij}}{2m}+V(q_{1}).  \tag{13}
\end{equation}%
It is evident from this expression that $H$ in (13) is not $SL(2,R)$%
-invariant Hamiltonian. Thus, a Hamiltonian of the form (12) does not admit
a $SL(2,R)$-invariant formulation. The same conclusion can be obtained by
considering a Hamiltonian constraint $H=\lambda (p^{i}p_{i}+m^{2})$ for the
relativistic point particle, where in this case $i$ runs from $0$ to $3$.

Thus, one finds that the simplest example of $SL(2,R)$-invariant Hamiltonian
seems to be [5]

\begin{equation}
H=\frac{1}{2}\lambda ^{ab}q_{a}^{i}q_{b}^{j}\eta _{ij},  \tag{14}
\end{equation}%
which can be understood as the Hamiltonian associated with a relativistic
harmonic oscillator in a phase space. Here, we assume that $\lambda
^{ab}=\lambda ^{ba}$ is a set of Lagrange multipliers and $\eta
_{ij}=diag(-1,-1,...,-1,1,...,1)$. Note that we are considering a signature
of the form $n=t+s$, with $t$ time-like and $s$ space-like signature. The
reason for this general choice is that the $SL(2,R)$-symmetry requires
necessarily a target `spacetime' with either $t=1$ and $s=1$ or $t\geqq 2$
and $s\geqq 2$ as we shall prove below.

Using (14) one sees that (10) can be written in the form

\begin{equation}
S=\frac{1}{2}\int_{t_{i}}^{t_{f}}dt\left( J^{ab}\dot{q}_{a}^{i}q_{b}^{j}\eta
_{ij}-\lambda ^{ab}H_{ab}\right) ,  \tag{15}
\end{equation}%
where

\begin{equation}
H_{ab}=q_{a}^{i}q_{b}^{j}\eta _{ij}.  \tag{16}
\end{equation}%
Of course $H_{ab}\approx 0$ is the constraint of the theory. Observe that
the constraint $H_{ab}\approx 0$ is symmetric in the indices $a$ and $b$,
that is $H_{ab}=H_{ba}$.

Note that using the definitions (8) and (9) we can write the usual Poisson
bracket (6), for arbitrary functions $f(q,p)$ and $g(q,p)$ of the canonical
variables $q$ and $p$ as

\begin{equation}
\{f,g\}=J_{ab}\eta ^{ij}\frac{\partial f}{\partial q_{a}^{i}}\frac{\partial g%
}{\partial q_{b}^{j}}.  \tag{17}
\end{equation}%
Thus, from (17) one discovers the algebra

\begin{equation}
\{q_{a}^{i},q_{b}^{j}\}=J_{ab}\eta ^{ij},  \tag{18}
\end{equation}%
which is equivalent to (7).

Moreover, using (17) one finds that $H_{ab}$ satisfies the $SL(2,R)$-algebra%
\begin{equation}
\{H_{ab},H_{cd}\}=J_{ac}H_{bd}+J_{ad}H_{bc}+J_{bc}H_{ad}+J_{bd}H_{ac}, 
\tag{19}
\end{equation}%
which shows that $H_{ab}$ is a first class constraint. Explicitly, the
nonvanishing brackets of the algebra (19) can be decomposed as

\begin{equation}
\{H_{11},H_{22}\}=4H_{12},  \tag{20}
\end{equation}%
\begin{equation}
\{H_{11},H_{12}\}=2H_{11},  \tag{21}
\end{equation}%
and%
\begin{equation}
\{H_{12},H_{22}\}=2H_{22}.  \tag{22}
\end{equation}%
By writing $S_{3}=-\frac{1}{2}H_{12}$, $S_{1}=\frac{1}{4}(H_{11}+H_{22})$
and $S_{2}=\frac{1}{4}(H_{11}-H_{22})$ one finds that%
\begin{equation}
\{S_{1},S_{2}\}=S_{3},  \tag{23}
\end{equation}%
\begin{equation}
\{S_{3},S_{1}\}=S_{2},  \tag{24}
\end{equation}%
and

\begin{equation}
\{S_{2},S_{3}\}=-S_{1},  \tag{25}
\end{equation}%
which can be succinctly written as

\begin{equation}
\{S_{\mu },S_{\nu }\}=\epsilon _{\mu \nu }^{\qquad \alpha }S_{\alpha } 
\tag{26}
\end{equation}%
or

\begin{equation}
\{S_{\mu \nu },S_{\alpha \beta }\}=\eta _{\mu \alpha }S_{\nu \beta }-\eta
_{\mu \beta }S_{\nu \alpha }+\eta _{\nu \beta }S_{\mu \alpha }-\eta _{\nu
\alpha }S_{\mu \beta }.  \tag{27}
\end{equation}%
Here\ $\eta _{\mu \nu }=(-1,1,1),$ $S_{\mu \nu }=-S_{\nu \mu }$ and $S^{\mu
}=\frac{1}{2}\epsilon ^{\mu \nu \alpha }S_{\nu \alpha }$, with $\epsilon
^{123}=-1$ and $\epsilon _{123}=1$. This is one way to see that the algebra $%
sl(2,R)$ is equivalent to the algebra $so(1,2)$. Furthermore, the group $%
SL(2,R)$ is double cover of $SO(1,2)$.

All this developments are relevant for quantization. In this case, one
defines the Poisson brackets in classical phase space and then associate
operators $\hat{f}(\hat{q},\hat{p})$ and $\hat{g}(\hat{q},\hat{p})$ to the
functions $\ f(q,p)$ and $g(q,p)$. Without constraints, the transition from
classical to quantum mechanics is made by promoting the canonical
Hamiltonian $H_{c}$ as an operator $\hat{H}_{c}$ via the nonvanishing
commutator

\begin{equation}
\lbrack \hat{q}^{i},\hat{p}_{j}]=i\delta _{j}^{i},  \tag{28}
\end{equation}%
(with $\hbar =1$) obtained from the second bracket in (7), and by writing
the quantum formula

\begin{equation}
\hat{H}_{c}|\Psi \rangle =i\frac{\partial }{\partial t}|\Psi \rangle , 
\tag{29}
\end{equation}%
which determines the physical states $|\Psi \rangle $ (see Refs. [9]-[11]
for details). Here, the bracket $[\hat{A},\hat{B}]=\hat{A}\hat{B}-\hat{B}%
\hat{A}$ denotes the commutator. This is in agreement with the meaning of $%
\hat{H}_{c}$ as the generator of temporal evolution for operators in the
Hilbert space.

If we have a constrained Hamiltonian system characterized by $m$ first class
constraints $H_{A}$, one also imposes that the correspondent operators acts
on the physical states as $\hat{H}_{A}|\Psi \rangle =0$.

\bigskip \ 

\noindent \textbf{4. }$SL(2,R)$\textbf{-symplectic structure and the
(2+2)-signature}

\smallskip

\noindent Applying Noether's procedure to (15) one learns that the angular
momentum

\begin{equation}
L^{ij}=q^{i}p^{j}-q^{j}p^{i}  \tag{30}
\end{equation}%
or

\begin{equation}
L^{ij}=J^{ab}q_{a}^{i}q_{b}^{j}  \tag{31}
\end{equation}%
is a conserved dynamic variable. Using (7) and (30) one can show that $%
L^{ij} $ obeys Lorentz group algebra

\begin{equation}
\{L^{ij},L^{kl}\}=\eta ^{ik}L^{jl}-\eta ^{il}L^{jk}+\eta ^{jl}L^{ik}-\eta
^{jk}L^{il}.  \tag{32}
\end{equation}%
Alternatively, one can show that this result also follows from (18) and (31).

We are now ready to write and prove the following proposition:

\smallskip \ 

\textbf{Proposition: }Let\textbf{\ }$(t+s)$ be the signature of the flat
metric $\eta _{ij}$ associated with a phase space described with coordinates 
$q_{a}^{i}$ which determine the $SL(2,R)$-symplectic structure given by the
Poisson brackets (17). Then, only in the cases $t=1$ and $s=1$ or $t\geqq 2$
and $s\geqq 2$ there exist coordinates $q_{a}^{i}$ different from zero such
that

\begin{equation}
H_{ab}=0  \tag{33}
\end{equation}%
and%
\begin{equation}
L^{ij}\neq 0.  \tag{34}
\end{equation}

\smallskip \ 

\textbf{Proof}: Consider a $SL(2,R)$-symplectic structure as in (17). For
the $\eta _{ij}$-symbol we shall assume the general case of $(t+s)$%
-signature corresponding to $t$-time and $s$-space coordinates $q^{i}$.
First observe that explicitly, (33) yields

\begin{equation}
q^{i}q^{j}\eta _{ij}=0,  \tag{35}
\end{equation}

\begin{equation}
q^{i}p^{j}\eta _{ij}=0,  \tag{36}
\end{equation}%
and

\begin{equation}
p^{i}p^{j}\eta _{ij}=0.  \tag{37}
\end{equation}%
Of course a theory with $t=0$ and $s=0$ is vacuous, so we shall assume that $%
t\neq 0$ or $s\neq 0$. From (35) and (37) one finds that if $t=0$ and $s\neq
0$, that is if $\eta _{ij}$ is Euclidean, then $q^{i}=0$ and $p^{i}=0$. This
shows the need for at least one time-like dimension, that is $t>0$. Note
that one can multiply (35)-(37) by a minus sign. This changes the signature
of $\eta _{ij}$ from $t+s$ to $s+t$. This means that if one assumes $t\neq 0$
and $s=0$, it results that the theory should have at least one space-like
dimension, that is $s>0$. So putting together these two results we have a
consistent solution of (35)-(37) only if $t\geqq 1$ and $s\geqq 1$.

We shall show that the case $t=1$ and $s=1$ is an exceptional solution of
(35)-(37). In this case, these expressions become

\begin{equation}
-(q^{1})^{2}+(q^{2})^{2}=0,  \tag{38}
\end{equation}

\begin{equation}
-q^{1}p^{1}+q^{2}p^{2}=0  \tag{39}
\end{equation}%
and

\begin{equation}
-(p^{1})^{2}+(p^{2})^{2}=0,  \tag{40}
\end{equation}%
respectively. Using (38) and (40), one can verify that (39) is an identity.
Thus (38) and (40) do not lead to any relation between $q$ and $p$ and
therefore in this case the angular momentum condition (34) is satisfied.

It remains to explore consistency when $t=1$ and $s\geqq 2$ (or $t\geqq 2$
and $s=1$ due to the sign freedom in (35)-(37)). A well known result is that
when $t=1$ and $s\geqq 2$ two light-like orthogonal vectors are necessarily
parallel. Hence, in this case we get the expression $q^{i}=ap^{i}$ which,
according to (30), implies $L^{ij}=0$. This clearly contradicts our
assumption (34). The same result holds for the case $t\geqq 2$ and $s=1$.
Hence, we have shown that (33) and (34) makes sense only if $t=1$ and $s=1$
or $t\geqq 2$ and $s\geqq 2$.

Therefore, since (34) is linked to the $SO(t,s)$-symmetry one may concludes
a consistent $SL(2,R)$-theory can be obtained only in the cases $SO(1,1)$ or 
$SO(t\geqq 2,s\geqq 2)$. From the perspective that $SO(2,2)$ is a minimal
alternative, we have shown that the signatures $(1+1)$ and $(2+2)$ are
exceptional.

An alternative method for arriving at the same result is as follows. Let us
separate from (35)-(37) one time variable in the form

\begin{equation}
-(q^{1})^{2}+q^{i^{\prime }}q^{j^{\prime }}\eta _{i^{\prime }j^{\prime }}=0,
\tag{41}
\end{equation}

\begin{equation}
-q^{1}p^{1}+q^{i^{\prime }}p^{j^{\prime }}\eta _{i^{\prime }j^{\prime }}=0, 
\tag{42}
\end{equation}%
and

\begin{equation}
-(p^{1})+p^{i^{\prime }}p^{j^{\prime }}\eta _{i^{\prime }j^{\prime }}=0, 
\tag{43}
\end{equation}%
where the indices $i^{\prime },j^{\prime }$, etc. run from $2$ to $t+s$. The
formula (42) leads to

\begin{equation}
(q^{1})^{2}(p^{1})^{2}-q^{i^{\prime }}p^{j^{\prime }}\eta _{i^{\prime
}j^{\prime }}q^{k^{\prime }}p^{l^{\prime }}\eta _{k^{\prime }l^{\prime }}=0.
\tag{44}
\end{equation}%
Using (41) and (43) we find that (44) becomes

\begin{equation}
q^{i^{\prime }}q^{j^{\prime }}\eta _{i^{\prime }j^{\prime }}p^{k^{\prime
}}p^{l^{\prime }}\eta _{k^{\prime }l^{\prime }}-q^{i^{\prime }}p^{j^{\prime
}}\eta _{i^{\prime }j^{\prime }}q^{k^{\prime }}p^{l^{\prime }}\eta
_{k^{\prime }l^{\prime }}=0,  \tag{45}
\end{equation}%
which can also be written as

\begin{equation}
(\delta _{i^{\prime }}^{j^{\prime }}\delta _{k^{\prime }}^{l^{\prime
}}-\delta _{i^{\prime }}^{l^{\prime }}\delta _{k^{\prime }}^{j^{\prime
}})q^{i^{\prime }}q_{j^{\prime }}p^{k^{\prime }}p_{l^{\prime }}=0.  \tag{46}
\end{equation}%
Observe that this implies that $\frac{1}{2}L^{i^{\prime }j^{\prime
}}L_{i^{\prime }j^{\prime }}=0$. If $\eta _{k^{\prime }l^{\prime }}$ is a
Euclidean metric this result in turn implies $L^{i^{\prime }j^{\prime }}=0$
which means that $q^{i^{\prime }}=\varsigma p^{i^{\prime }}$, that is $%
q^{i^{\prime }}$ and $p^{i^{\prime }}$ are parallel quantities. The
combination of (41) and (43) implies that $q^{1}=\varsigma p^{1}$. This is
another way to show that two light-like orthogonal vectors are parallel.

Let us now introduce the completely antisymmetric symbol

\begin{equation}
\varepsilon ^{i_{2}^{\prime }...i_{t+s}^{\prime }}.  \tag{47}
\end{equation}%
This is a rank-$t+s-1$ tensor which values are $+1$ or $-1$ depending on
even or odd permutations of%
\begin{equation}
\varepsilon ^{2...t+s},  \tag{48}
\end{equation}%
respectively. Moreover, $\varepsilon ^{i_{2}^{\prime }...i_{t+s}^{\prime }}$
takes the value $0$, unless the indices $i_{2}^{\prime }...i_{t+s}^{\prime }$
are all different.

Relation (46) can be written in terms of $\varepsilon ^{i_{2}^{\prime
}...i_{t+s}^{\prime }}$ in the form

\begin{equation}
\varepsilon ^{j^{\prime }l^{\prime }i_{4}^{\prime }...i_{t+s}^{\prime
}}\varepsilon _{i^{\prime }k^{\prime }i_{4}^{\prime }...i_{t+s}^{\prime
}}q^{i^{\prime }}q_{j^{\prime }}p^{k^{\prime }}p_{l^{\prime }}=0,  \tag{49}
\end{equation}%
where we have dropped the nonzero factor $\frac{1}{(t+s-2)!}$. Moreover,
(49) can be rewritten as

\begin{equation}
\varepsilon ^{j^{\prime }l^{\prime }i_{4}^{\prime }...i_{t+s}^{\prime
}}\varepsilon _{i^{\prime }k^{\prime }i_{4}^{\prime }...i_{t+s}^{\prime
}}L^{i^{\prime }k^{\prime }}L_{j^{\prime }l^{\prime }}=0.  \tag{50}
\end{equation}%
Here, we used (30) and dropped some numerical factors. Observe that 
\begin{equation}
L_{i_{4}^{\prime }...i_{t+s}^{\prime }}=\frac{1}{2}\varepsilon _{i^{\prime
}k^{\prime }i_{4}^{\prime }...i_{t+s}^{\prime }}L^{i^{\prime }k^{\prime }} 
\tag{51}
\end{equation}%
is the dual tensor of $L^{i^{\prime }k^{\prime }}$.

The lower dimensional case in which (50) holds is

\begin{equation}
\varepsilon ^{j^{\prime }l^{\prime }}\varepsilon _{i^{\prime }k^{\prime
}}L^{i^{\prime }k^{\prime }}L_{j^{\prime }l^{\prime }}=0,  \tag{52}
\end{equation}%
which implies

\begin{equation}
\varepsilon _{i^{\prime }k^{\prime }}L^{i^{\prime }k^{\prime }}=0.  \tag{53}
\end{equation}%
Consequently, this gives $L_{j^{\prime }l^{\prime }}=0$. Hence this proves
that the signature solutions $(1+2)$ or $(2+1)$ are not consistent with
(34). So, it remains to prove that $(1+(s>2))$ is also no consistent with
(34). In general we have that (50) and (51) imply%
\begin{equation}
L_{i_{4}^{\prime }...i_{t+s}^{\prime }}L^{i_{4}^{\prime }...i_{t+s}^{\prime
}}=0.  \tag{54}
\end{equation}%
But in the case $(1+s>2)$, (54) is an Euclidean expression and therefore $%
L_{i_{4}^{\prime }...i_{t+s}^{\prime }}=0$, which in turn implies $%
L_{j^{\prime }l^{\prime }}=0$. Thus, a consistent solution is also possible
in the case $t\geqq 2$ and $s\geqq 2$. Hence, this is an alternative proof
that with two time-like dimensions, the minimal case in which the $SL(2,R)$%
-symmetry is consistent with Lorentz symmetry, is the $2+2$-signature. In
principle we may continue with this procedure founding that $3+3$ and so on
are consistent possibilities. But, considering that (35)-(37) are only three
constraints we see that there are not enough constraints to eliminate all
additional degrees of freedom in all possible cases with $t\geq 3$ and $%
s\geq 3.$ In fact, one should expect that this will lead to unwanted results
at the quantum level [5]-[7].

Note what happens with the Lorentz Casimir operator

\begin{equation}
C\equiv \frac{1}{2}L^{ij}L_{ij}=\det (H_{ab}).  \tag{55}
\end{equation}%
From (31) we have

\begin{equation}
\begin{array}{c}
C=\frac{1}{2}L^{ij}L_{ij}=\frac{1}{2}%
J^{ab}q_{a}^{i}q_{b}^{j}J^{cd}q_{ci}q_{dj} \\ 
\\ 
=\frac{1}{2}J^{ab}J^{cd}q_{a}^{i}q_{ci}q_{b}^{j}q_{dj}=\frac{1}{2}%
J^{ab}J^{cd}H_{ac}H_{bd}.%
\end{array}
\tag{56}
\end{equation}%
Hence, when $H_{ab}=0$ we have $C=0$ which means that in this case the
Lorentz Casimir operator vanishes.

Summarizing, by imposing the $SL(2,R)$-symmetry and the Lorentz symmetry $%
SO(t,s)$ in the Lagrangian (15) we have shown that there exist $q_{a}^{i}$
consistent with these symmetries only in the signatures $1+1$ and $t\geqq
2+s\geqq 2$.

\bigskip \ 

\noindent \textbf{5.- The Dirac equation and the (2+2)-signature }

\smallskip \ 

\noindent As an application of our previous developments, in this section we
consider the Dirac equation in (2+2)-dimensions. This type of equation has
already be mentioned in [15], but here we construct it from first
principles. For this purpose, let us consider a relativistic point particle
described by the action%
\begin{equation}
S=-m_{0}\int d\tau \left( -\dot{x}^{\mu }\dot{x}^{\nu }\xi _{\mu \nu
}\right) ^{1/2}.  \tag{57}
\end{equation}%
In this section, we also use the notation $\dot{x}^{\mu }=\frac{dx^{\mu
}(\tau )}{d\tau }$, where $\tau $ is an arbitrary parameter. The tensor $\xi
_{\mu \nu }$ is a flat metric with signature $\xi _{\mu \nu
}=diag(-1,-1,1,1) $.

Starting from the Lagrangian associated with (57)

\begin{equation}
\mathcal{L}_{1}=-m_{0}\left( -\dot{x}^{\mu }\dot{x}^{\nu }\xi _{\mu \nu
}\right) ^{1/2},  \tag{58}
\end{equation}%
one finds that the canonical moments associated with $x^{\mu }$, namely

\begin{equation}
P_{\mu }={\frac{\partial {\mathcal{L}_{1}}}{\partial \dot{x}^{\mu }}}\text{,}
\tag{59}
\end{equation}%
lead to%
\begin{equation}
P_{\mu }{=}\frac{m_{0}\dot{x}^{\nu }\xi _{\mu \nu }}{\left( -\dot{x}^{\alpha
}\dot{x}^{\beta }\xi _{\alpha \beta }\right) ^{1/2}}{.}  \tag{60}
\end{equation}%
From (60), one can verify that

\begin{equation}
\mathbf{\mathcal{H}}\equiv P_{\mu }P_{\nu }\xi ^{\mu \nu }+m_{0}^{2}=0, 
\tag{61}
\end{equation}%
where $\xi ^{\mu \nu }=diag(-1,-1,1,1)$ is the inverse flat metric of $\xi
_{\mu \nu }$. Moreover, if we define the canonical Hamiltonian%
\begin{equation}
\mathbf{\mathcal{H}}_{\mathbf{c}}\equiv \dot{x}^{\mu }P_{\mu }-\mathcal{L}%
_{1},  \tag{62}
\end{equation}%
one sees that (60) also implies that

\begin{equation}
\mathbf{\mathcal{H}}_{\mathbf{c}}\equiv 0.  \tag{63}
\end{equation}

According to the Dirac constraint Hamiltonian system formalism, one can
write the total Hamiltonian as

\begin{equation}
\mathbf{\mathcal{H}}_{T}=\mathbf{\mathcal{H}}_{\mathbf{c}}+\lambda \mathbf{%
\mathcal{H}},  \tag{64}
\end{equation}%
where $\lambda $ is a Lagrange multiplier. By using the constraint (61), as
well as (63) and (64), one can write the first-order Lagrangian%
\begin{equation}
\mathcal{L}_{2}=\dot{x}^{\mu }P_{\mu }-\frac{\lambda }{2}(P_{\mu }P_{\nu
}\xi ^{\mu \nu }+m_{0}^{2}).  \tag{65}
\end{equation}

At the quantum level one requires to apply the constraint (61) to the
physical sates $\Phi $ in the form

\begin{equation}
\lbrack \hat{P}_{\mu }\hat{P}_{\nu }\xi ^{\mu \nu }+m_{0}^{2}]\Phi =0, 
\tag{66}
\end{equation}%
where $\hat{P}_{\mu }$ is an operator associated with $P_{\mu }$.

By starting with (66), our goal now is to construct a Dirac-type equation in 
$(2+2)$-dimensions. Let us first write (66) in the form%
\begin{equation}
\lbrack -\hat{P}_{1}\hat{P}_{1}+\hat{P}_{a}\hat{P}_{b}\eta
^{ab}+m_{0}^{2}]\Phi =0.  \tag{67}
\end{equation}%
Here, the flat metric $\eta ^{ab}$ is given by $\eta ^{ab}=diag(-1,1,1)$,
and the indices $a,b,...$ take values in the set $\{2,3,4\}$. Consider
matrices $\varrho ^{a}$ such that

\begin{equation}
\varrho ^{a}\varrho ^{b}+\varrho ^{b}\varrho ^{a}=2\eta ^{ab}.  \tag{68}
\end{equation}%
Using (68) one sees that (67) can be written as%
\begin{equation}
\lbrack (-\hat{P}_{1}+\varrho ^{a}\hat{P}_{a})(\hat{P}_{1}+\varrho ^{b}\hat{P%
}_{b})+m_{0}^{2}]\Phi =0.  \tag{69}
\end{equation}%
Now, we define two spinors

\begin{equation}
\Phi _{L}\equiv \Phi  \tag{70}
\end{equation}%
and

\begin{equation}
\Phi _{R}\equiv -\frac{1}{m_{0}}(\hat{P}_{1}+\varrho ^{b}\hat{P}_{b})\Phi
_{L}.  \tag{71}
\end{equation}%
Explicitly (71) leads to%
\begin{equation}
(\hat{P}_{1}+\varrho ^{b}\hat{P}_{b})\Phi _{L}+m_{0}\Phi _{R}=0,  \tag{72}
\end{equation}%
while (69), (70) and (71) give%
\begin{equation}
(\hat{P}_{1}-\varrho ^{a}\hat{P}_{a})\Phi _{R}+m_{0}\Phi _{L}=0.  \tag{73}
\end{equation}%
These last two equations can be expressed in a matrix form%
\begin{equation}
\left( 
\begin{bmatrix}
0 & I \\ 
I & 0%
\end{bmatrix}%
\hat{P}_{1}+%
\begin{bmatrix}
0 & \varrho ^{a} \\ 
-\varrho ^{a} & 0%
\end{bmatrix}%
\hat{P}_{a}+%
\begin{bmatrix}
I & 0 \\ 
0 & I%
\end{bmatrix}%
m_{0}\right) \left( 
\begin{array}{c}
\Phi _{R} \\ 
\Phi _{L}%
\end{array}%
\right) =0,  \tag{74}
\end{equation}%
where $I=diag(1,1)$ is the identity matrix in two dimensions. One can of
course write (74) in the more compact form%
\begin{equation}
(\Gamma ^{\mu }\hat{P}_{\mu }+m_{0})\Psi =0.  \tag{75}
\end{equation}%
Here, we used the following definitions

\begin{equation}
\Psi \equiv \left( 
\begin{array}{c}
\Phi _{R} \\ 
\Phi _{L}%
\end{array}%
\right) ,  \tag{76}
\end{equation}%
\begin{equation}
\Gamma ^{1}\equiv 
\begin{bmatrix}
0 & I \\ 
I & 0%
\end{bmatrix}%
,  \tag{77}
\end{equation}%
and%
\begin{equation}
\Gamma ^{a}\equiv 
\begin{bmatrix}
0 & \varrho ^{a} \\ 
-\varrho ^{a} & 0%
\end{bmatrix}%
.  \tag{78}
\end{equation}%
By promoting $\hat{P}_{\mu }\rightarrow i\partial _{\mu }$, one recognize in
(75) the Dirac type equation in $(2+2)$-dimensions.

We shall show that (75) is deeply linked to the $SL(2,%
\mathbb{R}
)$-group. First, observe that an explicit representation of the matrices $%
\varrho _{1}$ and $\varrho _{a}$ in (78) is%
\begin{equation}
\begin{array}{cc}
\varrho _{1}=%
\begin{pmatrix}
1 & 0 \\ 
0 & 1%
\end{pmatrix}%
, & \varrho _{2}=%
\begin{pmatrix}
0 & -1 \\ 
1 & 0%
\end{pmatrix}%
, \\ 
\varrho _{3}=%
\begin{pmatrix}
1 & 0 \\ 
0 & -1%
\end{pmatrix}%
, & \varrho _{4}=%
\begin{pmatrix}
0 & 1 \\ 
1 & 0%
\end{pmatrix}%
.%
\end{array}
\tag{79}
\end{equation}%
Notice first that the determinant of each of the matrices (79) is different
from $0$. This suggests to relate such matrices with the general group $GL(2,%
\mathbb{R}
)$. Indeed, the matrices in (79) can be considered as a basis for a general
matrix $M$ in the following manner:

\begin{equation}
M=%
\begin{pmatrix}
A & B \\ 
C & D%
\end{pmatrix}%
=\varrho _{1}a+\varrho _{2}b+\varrho _{3}c+\varrho _{4}d,  \tag{80}
\end{equation}%
where $a,b,c,d\in 
\mathbb{R}
$, given by%
\begin{equation}
\begin{array}{cc}
a=\frac{1}{2}(A+D), & b=\frac{1}{2}(-B+C), \\ 
&  \\ 
c=\frac{1}{2}(A-D), & d=\frac{1}{2}(B+C),%
\end{array}
\tag{81}
\end{equation}%
Explicitly, (80) can be read

\begin{equation}
M=%
\begin{pmatrix}
a+c & -b+d \\ 
b+d & a-c%
\end{pmatrix}%
.  \tag{82}
\end{equation}%
Without loss of generality, one may assume that $\det (M)\neq 0$, in such a
way that $M$ is contained in the Lie group $GL(2,%
\mathbb{R}
)$. If one also impose the condition that $\det (M)=1$, the matrix $M$
belongs to the Lie group $SL(2,%
\mathbb{R}
)$.

It is worthwhile to mention that, by writing $\varrho _{a}$ in tensorial
notation%
\begin{equation}
\begin{array}{ccc}
\varepsilon _{ij}=\varrho _{2}, & \eta _{ij}=\varrho _{3}, & \lambda
_{ij}=\varrho _{4},%
\end{array}
\tag{83}
\end{equation}%
one can construct a gravity model in 2 dimensions (see Ref. [16] for
details).

Rewriting (72) and (73) respectively as follows

\begin{equation}
(\varrho _{1}\hat{P}_{1}+\varrho _{2}\hat{P}_{2}+\varrho _{3}\hat{P}%
_{3}+\varrho _{4}\hat{P}_{4})\Phi _{L}+m_{0}\Phi _{R}=0,  \tag{84}
\end{equation}%
and%
\begin{equation}
(\varrho _{1}\hat{P}_{1}-\varrho _{2}\hat{P}_{2}-\varrho _{3}\hat{P}%
_{3}-\varrho _{4}\hat{P}_{4})\Phi _{R}+m_{0}\Phi _{L}=0,  \tag{85}
\end{equation}%
one sees that both (84) and (85) have the matrix form (80). This means that
these two equations can be indentified with the Lie group $SL(2,%
\mathbb{R}
)$. Indeed, taking into account (80), we see that (84) and (85) can be
rewritten as

\begin{equation}
\begin{bmatrix}
\hat{P}_{1}+\hat{P}_{3} & -\hat{P}_{2}+\hat{P}_{4} \\ 
\hat{P}_{2}+\hat{P}_{4} & \hat{P}_{1}-\hat{P}_{3}%
\end{bmatrix}%
\Phi _{L}+m_{0}\Phi _{R}=0.  \tag{86}
\end{equation}%
and%
\begin{equation}
\begin{bmatrix}
\hat{P}_{1}-\hat{P}_{3} & \hat{P}_{2}-\hat{P}_{4} \\ 
-\hat{P}_{2}-\hat{P}_{4} & \hat{P}_{1}+\hat{P}_{3}%
\end{bmatrix}%
\Phi _{R}+m_{0}\Phi _{L}=0,  \tag{87}
\end{equation}%
respectively. One observes that (86) and (87) are matrix-like moments
similar to the general matrix (80). Similarly, one can identify the moments
matrices contained in the expressions (86) and (87) with the symmetry group $%
SL(2,%
\mathbb{R}
)$. Let us introduce a new momenta matrix%
\begin{equation}
\mathcal{\hat{P}}^{\pm }=\frac{1}{m_{0}}%
\begin{bmatrix}
\hat{P}_{1}\pm \hat{P}_{3} & \pm (-\hat{P}_{2}+\hat{P}_{4}) \\ 
\pm (\hat{P}_{2}+\hat{P}_{4}) & \hat{P}_{1}\mp \hat{P}_{3}%
\end{bmatrix}%
.  \tag{88}
\end{equation}%
Consequently, the equations (86) and (87) become%
\begin{equation}
\mathcal{\hat{P}}^{+}\Phi _{L}+\Phi _{R}=0  \tag{89}
\end{equation}%
and%
\begin{equation}
\mathcal{\hat{P}}^{-}\Phi _{R}+\Phi _{L}=0.  \tag{90}
\end{equation}%
Note that taking into account the constraint (86) we have

\begin{equation}
\det \mathcal{\hat{P}}^{\pm }\Phi _{R,L}=\Phi _{R,L}.  \tag{91}
\end{equation}%
Symbolically, we can consider 
\begin{equation}
\det \mathcal{\hat{P}}^{\pm }=I  \tag{92}
\end{equation}%
But this means that both $\mathcal{\hat{P}}^{+}$ and $\mathcal{\hat{P}}^{-}$
are elements of $SL(2,%
\mathbb{R}
)$-group and therefore the Dirac type equation (74) or (79) has a structure
associated with the group $SL(2,%
\mathbb{R}
)^{+}\times SL(2,%
\mathbb{R}
)^{-}$. In fact, this may be understood considering the isomorphism $%
SO(2,2)\sim SL(2,%
\mathbb{R}
)\times SL(2,%
\mathbb{R}
)$.

As it is known, the Dirac equation describes massive particles with $\frac{1%
}{2}$-spin. When the mass $m_{0}$ is the mass of the electron, the Dirac
equation correctly determines the quantum theory of the electron. On the
other hand, the Dirac type equation (74) in $(2+2)$-dimensions also
describes massive particles with $\frac{1}{2}$-spin,. However, there is a
significant distinction for this signature: while in the case of Dirac
equation in $(1+3)$-dimensions $\Psi $ can be choosen as a Majorana or Weyl
spinor (but not both at the same time), one can choose $\Psi $ as a
Majorana-Weyl spinor in $(2+2)$-dimensions.

\smallskip \ 

\noindent \textbf{6. Final Comments}

\smallskip

\noindent We have proved in some detail that $SL(2,R)$-symmetry and Lorentz
symmetry $SO(t,s)$ imply together that the signatures $1+1$ and $2+2$ are
exceptional. One may be motivated to relate this result with different
physical scenarios. Of course, the signature $1+1$ can be related to string
theory. But what about the $2+2$ signature? We already know that this
signature arises in a number of physical scenarious, including in a
background for $N=2$ strings [17]-[18] (see also Refs [19]-[21]), Yang-Mills
in Atiyah Singer background [22] (see also Refs. [23] for the importance of
the $2+2$ signature in mathematics), Majorana-Weyl spinor [24]-[25] and more
recently in loop quantum gravity in terms of oriented matroid theory [26]
(see also references [27]-[29]). But one wonders whether the $2+2$ signature
can be linked to quantum gravity itself in $1+3$ dimensions. One possibility
to answer this question is to search for a mechanism which can transform
self-dual canonical gravity in $2+2$ dimensions into self-dual canonical
gravity in $1+3$. This is equivalent to change one time dimension by one
space dimension and \textit{vice versa}. Surprisingly this kind of
transformation has already be considered in the context of the sigma model
(see Ref. [30] and references therein). In fact, it was shown in [27] that
similar mechanism can be implemented at the level of quantum self-dual
canonical gravity $2+2$ dimensions.

\bigskip

\begin{center}
\textbf{Acknowledgments}
\end{center}

This work was partially supported by PROFAPI-UAS 2009.

\end{document}